\documentclass[sigconf]{acmart}

\usepackage{booktabs} 
\usepackage[colorinlistoftodos]{todonotes}

\usepackage{color,soul}

\setcopyright{rightsretained}

\acmDOI{10.475/123_4}

\acmISBN{123-4567-24-567/08/06}

\acmConference[MET 2018]{International Workshop on Metamorphic Testing}{May 2018}{Gothenburg, Sweden} 
\acmYear{2018}
\copyrightyear{2018}

\acmArticle{4}
\acmPrice{15.00}

\begin{document}
\title{Predicting Metamorphic Relation for Matrix Calculation Programs}


\author{Karishma Rahman}
\orcid{1234-5678-9012}
\affiliation{%
  \institution{Montana State University}
  \city{Bozeman} 
  \state{Montana}
}
\email{karishma.rahman@student.montana.edu}

\author{Upulee Kanewala}
\authornote{Corresponding author}
\affiliation{%
  \institution{Montana State University}
  \city{Bozeman} 
  \state{Montana} 
}
\email{upulee.kanewala@montana.edu}

\renewcommand{\shortauthors}{K. Rahman and U. Kanewala}

\begin{abstract}
Matrices often represent important information in scientific applications and are involved in performing complex calculations. But systematically testing these applications is hard due to the oracle problem. Metamorphic testing is an effective approach to test such applications because it uses metamorphic relations to determine whether test cases have passed or failed. Metamorphic relations are typically identified with the help of a domain expert and is a labor intensive task. In this work we use a graph kernel based machine learning approach to predict metamorphic relations for matrix calculation programs. Previously, this graph kernel based machine learning approach was used to successfully predict metamorphic relations for programs that perform numerical calculations. Results of this study show that this approach can be used to predict metamorphic relations for matrix calculation programs as well.   

\end{abstract}

%
%


\maketitle

\section{Introduction}
 \textit{Matrix calculations} are common in scientific applications. Often, matrices represent data, graphs or mathematical equations in the applications. \cite{office_2013}. They can be used to get quick and good approximation for complicated calculation in time-sensitive engineering applications \cite{office_2013}. Moreover, matrix multiplication is used in graphics, digital videos and solving linear equations of particular variables in different applications \cite{office_2013}. But testing these applications is hard due to the difficulties associated with defining suitable test oracles \cite{ref2}. This is known as the oracle problem \cite{ref2}. 
\textit{Metamorphic Testing (MT)} can be used to alleviate the test oracle problem \cite{ref3}. MT conducts testing by checking whether the programs behave according to a set of \textit{metamorphic relation (MR)} properties \cite{CHEN20031}. A metamorphic relation specifies how the output should change according to a change made to the input \cite{CHEN20031}. MT operates as follows \cite{ref3,CHEN20031}: 
\begin{enumerate}
\item Identify a suitable set of metamorphic relations which should satisfy the program under test. \item Create a set of initial test cases.
\item Apply the input transformations specified by the identified MRs in Step 1 and create follow-up test cases for each of the initial test case.
\item Execute the initial and follow-up test case pairs and check if the output change satisfies the change predicted by the MR. When testing a program, a run-time violation of an MR can mean that a fault or faults are present in the program under test.
\end{enumerate}
In a previous work \cite{ref4}, a \textit{graph kernel-based machine learning} method was introduced to predict MRs for programs with numerical inputs and outputs. In this work, we use the above method to predict MRs for functions performing matrix calculations. This method starts by creating the \textit{control flow graphs (CFGs)} of each program, and the random walk kernel is used to compute the similarity between the graphs. The computed kernel values are used by a \textit{support vector machine (SVM)} to automatically predict MRs for previously unseen functions. In this study, three types of metamorphic relations are identified for the matrix-based programs and are used for the predictions. We used 55 functions obtained from open source matrix calculation libraries to evaluate the effectiveness of this method. Our result shows that for matrix-based calculations, the random walk kernel can effectively predict the MRs. \par
\begin{table*}

  \caption{The Metamorphic Relations used in the study}
  \label{tab1}
  \begin{tabular}{lll}
    \toprule
    Metamorphic Relation&Change made to the input&Expected change in the output\\
    \midrule
   &Permutation of all the elements&\\
   Permutative&Permutation of rows&The matrix size will remain same \\
   &Permutation of columns& \\ \midrule
   &Scalar addition to matrix& \\
   Additive&Addition of two or more matrices&Element values will increase or remain same \\
   &Addition to the subset of elements of the matrix & \\
  \midrule
    Multiplicative&Scalar multiplication to matrix&Element values will increase\\
   &Multiplication to the subset of  elements of the matrix& \\
  \bottomrule
\end{tabular}
\end{table*}

\section{Approach}
\label{sec:method}
This section discusses the details of the metamorphic relation approach used in this study. 
\subsection{Function Representation}
The first step of this method is to convert a function into its CFG. This representation is specifically used since it allows the extraction of information about the sequence of operations performed in a control flow path that is directly related to the MRs satisfied by a given function. 

A CFG is a directed graph \(G\textsubscript{\textit{f}} = (V,E)\) of a function \textit{f}. Here, \textit{x} is a statement in \textit{f}, represented by each node \(v\textsubscript{\textit{x}} \in V \). The operation performed in each \textit{x} are labeled \textit{label}(\(v\textsubscript{x}\)). Supposedly if \textit{x} and \textit{y} are statements of \textit{f}, after execution of \textit{x}, \textit{y} is executed. Then it can be said that \textit{e} is an edge where \(e=(v\textsubscript{\textit{x}},v\textsubscript{\textit{y}}) \in E\). Control flow of the function \textit{f} is represented by all the edges, and the starting point and the exiting point are represented by nodes \(v\textsubscript{\textit{start}}\) and \(v\textsubscript{\textit{exit}}\) respectively~\cite{ref5}. 

\begin{figure}
\includegraphics[width=0.5\textwidth]{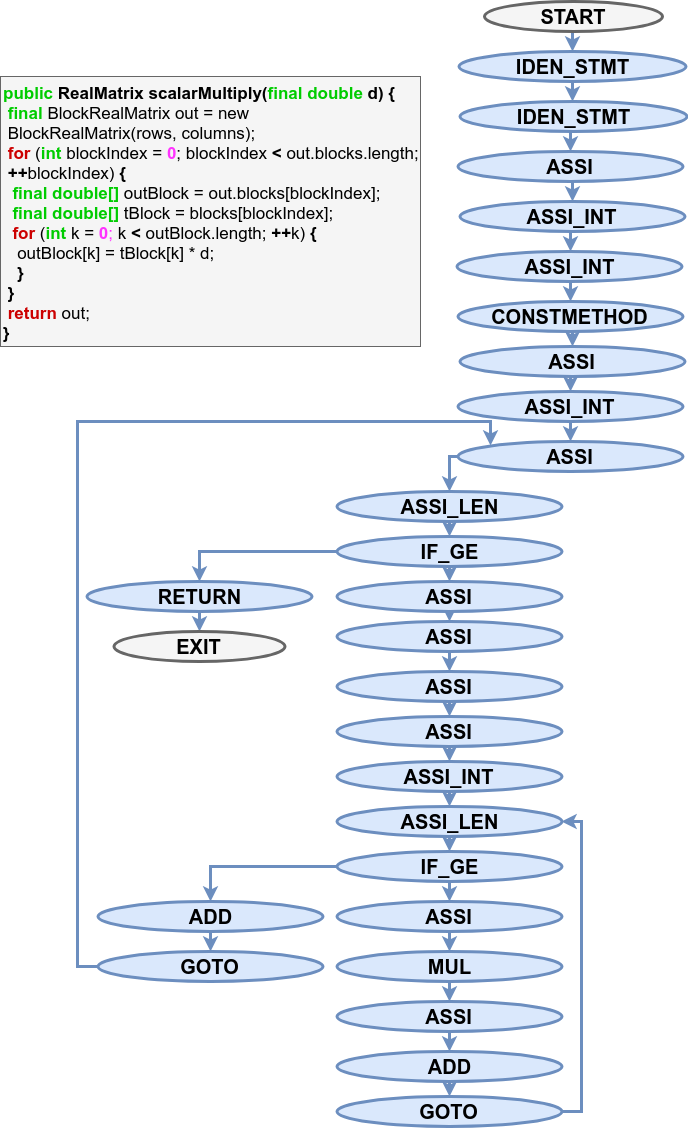}
\caption{Scalar multiply function and its post-processed CFG representation}
\label{figcfg}
\end{figure}

We use the \textit{Soot}\footnote{https://www.sable.mcgill.ca/soot/} framework to create the CFGs. We post-processed the generated CFGs from Soot so that the nodes would represent atomic operations. In addition we annotated all the method call nodes in the CFG with their return types. Figure \ref{figcfg} represents a function for calculating scalar multiplication of a matrix and its post-processed CFG representation.

\subsection{Random Walk Kernel}
After creating the CFG representation of the functions, the next step is to use a graph kernel to compute the similarity between the CFGs. In previous work \cite{ref4}, two graph kernels were used and among them, better performance was shown by the random walk kernel. Therefore we use the random walk kernel in this study. We briefly describe the idea of the random walk kernel in this section. More information about this including the definitions can be found in \cite{ref4}.

The random walk kernel computes the similarity score between two graphs by summing up the similarity scores of all the pairs of walks in the two graphs. The similarity score of a pair of walks is computed by multiplying the similarity scores of their corresponding step pairs. The similarity score of a pair of steps is computed by multiplying the similarity scores of node and edge pairs that make up the step. The similarity score of a node pair is determined by their node labels: if the two node labels are the same, then the pair is assigned a similarity score of one, else it is assigned a similarity score of zero. Also, if the two node labels represent operations with similar mathematical properties (but not identical), then the pair is assigned a similarity score of 0.5. Edge labels decide the value assigned for the similarity score of a pair of edges. In this work we only used one type of edge showing the flow of control between the operations. Thus the similarity score for a pair of edges is always one.

\subsection{Predictive Model Creation}
The computed random walk kernel values are supplied to a support vector machine with a binary label indicating whether a given function satisfies a given MR or not. The support vector machine uses the provided information to create a model that can predict if a new function would satisfy the considered MR or not. In this study, the SVM implementation from the scikit-learn\footnote{http://scikit-learn.org/stable/} toolkit was used.

\section{Experimental setup}
This section describes the code corpus and  MRs used in this study. The details of the evaluation procedure are also discussed here.
\subsection{The Code Corpus}
A total of 55 functions, all of which takes matrices as inputs and produces matrices as outputs, were used to measure the effectiveness of the method described in Section~\ref{sec:method} for predicting MRs. These functions were collected from \textit{Apache Commons Math Library}\footnote{https://commons.apache.org/proper/commons-math}, which is an open source project. 
These functions execute a variety of calculations on matrices such as addition, multiplication, subtraction, and searching (e.g. getting column matrix, getting row matrix).
There were several functions that performed the same functionality, but they were implemented differently. For example, Array2DRowRealMatrix class and OpenMapRealMatrix class both have multiplication functions for matrices, but they are implemented in different ways. In such cases, both the functions are used in the code corpus. All the functions used in this study can be found via the following URL: \url{https://github.com/MSU-STLab/MRPrediction/tree/master/alldotfiles} 
\begin{table}
  \caption{Number of positive and negative instances for each metamorphic relation}
  \label{tab2}
  \begin{tabular}{lcc}
    \toprule
    Metamorphic Relation&Positive instances&Negative instances\\
    \midrule
   Permutative&14&41\\ \midrule
   Additive&37&18\\\midrule
   Multiplicative&21&34 \\
  \bottomrule
\end{tabular}
\end{table}
\subsection{Metamorphic Relations} 
We manually identified three categories of MRs - Additive, Permutative, and Multiplicative, that are generally applicable to matrix calculations. These three high-level categories are further divided based on whether the modification is made at the element, row, or column levels. The full categorization of the MRs is shown in Table \ref{tab1}. In this work we only focus on predicting the high level MR category; i.e. Permutative, Additive and Multiplicative.  
\subsection{Evaluation Procedure}
We use \textit{train, validation and test method} to evaluate the MR prediction effectiveness. Table \ref{tab2} shows the number of positive and negative instances for each MR; positive indicates that a function satisfies the given MR and negative indicates that the function does not satisfy the given MR. For each MR, we divided the data into three subsets,
where each fold contained approximately the same portion of positive and negative instances, as the original dataset. The three folds were named as Train data, Test data, and Validation data.
The precomputed kernel values of the functions in Train data were used to create the prediction model. The Validation data was used to select the following parameters for the predictive model:
\begin{itemize}
\item Regularization parameter $C$ of the SVM.
\item Path weighing factor $\lambda$ in the random walk kernel where $0 \leq\lambda<1$.
\end{itemize}

The parameter values selected using the validation set were then used to create the predictive model for predicting the MRs for the test data. We repeated the train, validation and test method ten times so that the functions in each fold is selected randomly each time to avoid any biases occur in fold divisions.

We used the Area Under the receiver operating characteristic Curve (AUC) \cite{1388242} to measure the prediction effectiveness. AUC measures the probability that a randomly chosen negative example will have a lower prediction score than a randomly chosen positive example. AUC does not depend on the discrimination threshold of the classifier and has been shown to be a better measure for comparing learning algorithms \cite{1388242}.
\begin{table}
  \caption{Best C and $\lambda$ parameter of train model for each metamorphic relation}
  \label{tab3}
  \begin{tabular}{lcc}
    \toprule
    Metamorphic Relation&Best $\lambda$&Best C\\
    \midrule
   Permutative&0.9&0.1, 1, 10, 100, 1000\\ \midrule
   Additive&0.9&0.1, 1, 10, 100, 1000\\\midrule
   Multiplicative&0.9&0.1, 1, 10, 100, 1000 \\
  \bottomrule
\end{tabular}
\end{table}
\begin{figure}
\includegraphics[width=0.5\textwidth]{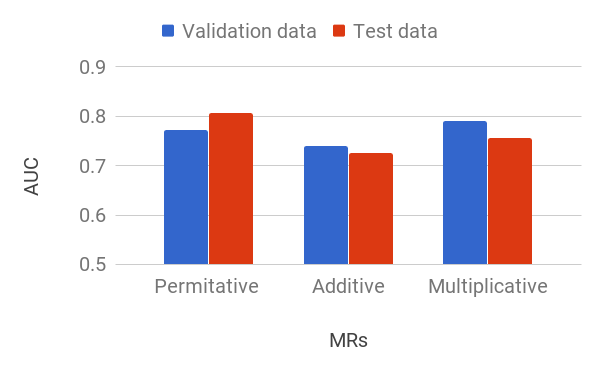}
\caption{Prediction \textit{AUC} score for \textit{random graph kernel} for \textit{validation dataset} and \textit{test dataset}}
\label{fig1}
\end{figure}
\section{Results and Discussion}

Table \ref{tab3} lists the $\lambda$ and C values that recorded the highest AUC values for each MR on the validation set. For the three MRs considered in this study, the value selected for the parameter $C$ doesn't seem to have a big effect on the prediction accuracy. But for all the three MRs, the best value for $\lambda$ is 0.9, indicating that longer paths in the CFGs are more important for predicting these MRs than the other paths.

Figure \ref{fig1} shows the AUC scores for the validation data set and the test data set. On the test data, the highest AUC score (0.81) could be observed when predicting the Permutative MR. The other two MRs also reported AUC values higher than 0.7 indicating that our approach created effective predictive models for all the three MRs. Further, for all the three MRs, AUC values for the validation data set and the test data set is close. This indicates that there is a low chance of over-fitting in the predictive model. 
\section{Related Work}
Several previous studies have looked into automatically generating/predicting MRs. Kanewala \textit{et. al} showed that, in previously unseen programs, MRs can be predicted using a machine learning method. Features were extracted from CFGs of the functions and they were then used to create a predictive model \cite{6698899}. Later, they developed the graph kernel based approach used in this study to predict MRs for numerical programs \cite{ref4}.\par 
Liu \textit{et al.} introduced a new method called \textit{Composition of Metamorphic Relation (CMR)}, where the generation of new metamorphic relations is done by combining existing metamorphic relations \cite{6319226}. A similar study has been done by Dong \textit{et. al}, where \textit{Compositional MR} was generated based on the speculative law of proposition logic \cite{dong}.\par
Zhang \textit{et al.} suggested a technique, where an algorithm searches for metamorphic relations in the form of linear or quadratic equations \cite{zhang11}. Su \textit{et al.} also suggested a new method called \textit{KABU}, which can be used to find more likely metamorphic relations by dynamically inferring the properties of the status of a method  \cite{Su:2015:DIL:2819261.2819279}.\par
Chen \textit{et al.} proposed a tool called \textit{METRIC}, where metamorphic relations were identified based on the category-choice framework \cite{metric}. Later, they introduced an approach called \textit{DESSERT}, where DividE-and-conquer methodology was used to identify the categorieS, choiceS, and choicE Relations for Test case generation \cite{5963695}. 
 \par
\section{Conclusion \&\ future work}
The metamorphic testing technique is very useful to test programs that do not have a test oracle. The effectiveness of this technique highly depends on the set of MRs used for testing. But the identification process of MRs is mostly done manually and could be a time consuming process.\par 
This study is an extension of previous work, where the random walk kernel is used to predict MRs for functions that performs matrix calculation.
Our results show that for these types of functions, random walk kernel can be effective in predicting MRs.\par
In the future, we plan to increase the number of functions used in this study. Further, new types of MRs, specifically for functions that perform matrix calculation, can also be considered. We also plan to extend the MR prediction scope beyond the function level. 

 \begin{acks}

This work is supported by award number 1656877 from the National Science Foundation. Any Opinions, findings and conclusions or recommendations expressed in this material are those of the author(s) and do not necessarily reflect those of the National Science Foundation.


\end{acks}

\bibliographystyle{ACM-Reference-Format}
\bibliography{sample-bibliography} 

\end{document}